\documentstyle[11pt]{article}
\input{epsfig.sty}

\newcommand{\cA}{{\cal A}}
\newcommand{\be}{\begin{equation}}
\newcommand{\ee}{\end{equation}}
\newcommand{\ba}{\begin{array}}
\newcommand{\ea}{\end{array}}
\newcommand{\alp}{\alpha}
\def\bq{\mbox{\boldmath $q$}}

\def\br{\mbox{\boldmath $r$}}

\begin{document}


\pagestyle{plain}

\title{ 
{\bf Nuclear matter 
symmetry energy
from generalized
polarizabilities: 
dependences on 
 momentum, isospin, density and temperature 
 } }
\author{  {\bf F\'abio L. Braghin} \thanks{e-mail: 
braghin@if.usp.br}  \\
{\normalsize 
Instituto de F\'\i sica, Universidade de 
S\~ao Paulo, C.P. 66.318; CEP 05315-970;  S\~ao Paulo - SP, Brazil. }\\
}
\date{}
\maketitle
\begin{abstract} 
Symmetry energy terms from macroscopic mass formulae
are investigated as generalized polarizabilities of nuclear matter. 
Besides the
neutron-proton (n-p) symmetry energy the spin dependent  
symmetry energies and a scalar one are also defined.
They depend on the nuclear densities ($\rho$), neutron-proton asymmetry
($b$),
temperature ($T$) and exchanged energy and momentum ($q$).
Based on a standard expression for the generalized polarizabilities, 
 a differential equation  is proposed to constrain
the dependence of the symmetry energy 
on the n-p asymmetry and
on the density. 
Some solutions are discussed.
The q-dependence (zero frequence) 
of the symmetry energy coefficients 
with Skyrme-type forces is investigated
in the four channels of the particle-hole interaction.
Spin dependent symmetry energies are also investigated 
indicating much stronger differences in behavior with $q$
for each Skyrme force than the results for the neutron-proton one. 
\end{abstract}

%

IF- USP - 2004

\section{Introduction}

The symmetry energy terms and their dependences on the
density are of relevance for the nuclear structure and
in many nuclear processes including
the structure and dynamics of proto-neutron and neutron stars.
The  neutron-proton
symmetry energy is the best known in spite of the
different values at the saturation density in the literature
(in the range of $25$MeV up to $36$MeV).
It is basically represented by a squared power of the 
neutron-proton (number or density) 
asymmetry in usual macroscopic/microscopic mass formula \cite{MONI},
the {\it parabolic approximation}. 
With a symmetry energy 
coefficient (s.e.c), $a_{\tau}$, 
the binding energy is, in the simplified versions,
 usually written as: 
\be \label{1}
\displaystyle{ E/A = H_0(A,Z)/A + a_{\tau}(N-Z)^2/A^2,}
\ee
where the energy density $H_0$  does not depend on the asymmetry,
Z, N and A are the proton, neutron and mass numbers respectively. 
Neutrons and protons occupying the same total volume yield
a term proportional to the squared asymmetry density, 
$a_{\tau}(\rho_N-\rho_Z)^2/\rho^2$ which appears in the nucleonic 
matter equation of state. 
The neutron and proton densities may not
 be exactly equal to each other in nuclei
\cite{NEUTRONdensity}.
Different polynomial terms of the asymmetry in this  expression
(proportional to $(N-Z)^n$ for $n \neq 2$)
are usually expected to be less relevant
\cite{MONI,HUBERetal,JACO,LKLB,NEEG,SATULAWYSS,DANIELEWICZ}.
However it is not well known whether and how this parabolic approximation
is to be modified for
very asymmetric systems,  such as nuclei far from the stability line
 or for (asymmetric) 
nuclear matter above and below the
saturation density \cite{EOS-rhodense}.
In large stable nuclei such as $^{208}Pb$ the n-p asymmetry
($(N-Z)^2/A^2 \simeq 1/9$)
is not so large as it would be in neutron matter.
The n-p symmetry energy coefficient (s.e.c.) $a_{\tau}$  is
given by the static polarizability of the system \cite{BVA}
(the inverse of the "isovector screening function")
which can also depend on the asymmetry of the medium \cite{NPA2000-2001}. 
This may lead to slightly different forms for
the symmetry energy for very asymmetric n-p systems.

Other symmetry energy coefficients may also be defined in 
nuclear matter, for instance, the
 spin $A_{\sigma}$ and  spin-isovector
$A_{\sigma \tau}$ ones. 
Extending  the n-p symmetry energy,
the other  symmetry energy coefficients
can be defined in  macroscopic mass formulae as:
\be \label{2} \ba{ll}
\displaystyle{ \frac{E}{A} = \frac{H_0(A,Z)}{\rho} + 
A_{\tau}\frac{(\rho_N- \rho_Z)^2}{\rho^2} +
A_{\sigma}\frac{( S_{up} - S_{down})^2}{(  S_{up} + S_{down} )^2} +
A_{\sigma \tau}\frac{( \rho^N_{up} - \rho^N_{down} + 
\rho^Z_{down} - \rho^Z_{up})^2}{
\rho^2}
,
}
\ea \ee
where the density (and eventually number) of 
 neutrons and protons is denoted by $\rho_N, \rho_Z$,  of 
nucleons 
with spin up (down) by $S_{up} (S_{down})$ and
$\rho_{up, down}^i$ the neutron/proton densities with
spin (up, down).
The spin channel may lead to
the appearance of polarized nucleonic matter which
has been investigated within different approaches with controversial
results
\cite{KUTSCHERAW1,KAISER,VIDAURREetal84,MARCOSetal91,BERNARDOSetal95,NPA2000-2001,FSS2001,VPR,VB02,ISAYEV,BJP-IJMPE}.
The spin channel is also relevant 
for the study of the neutrino interaction
with matter because it couples to the 
axial vector current together with 
the scalar channel in dense stars
\cite{SAWYER,ESPANHOIS,REDDY}.
The spin-isospin channel has been associated to
pion condensation  \cite{MIGDAL,AKMAL-PANDHAR}
and also  to anti-ferromagnetic states \cite{ISAYEV}.
A nuclear dipolar incompressibility was also defined 
in \cite{NPA2000-2001},
being related to the nuclear matter incompressibility as discussed 
below,
and which varies accordingly with the n-p asymmetry
being eventually relevant for 
the isoscalar dipole resonances \cite{ISDGR-YOUNGBLOODetal}.
These coefficients  and their corresponding
dependences on the asymmetry of neutron-proton densities
have been   investigated in several other works, as for example in
 \cite{YOSHIDA,PAL,BJP-IJMPE}.
A different way of obtaining the symmetry energy has been 
proposed by means of
the linear response method for the dynamical polarizabilities.
The static limit of these generalized polarizabilities are 
proportional to the inverse of the
symmetry energy coefficients in symmetric
matter \cite{BVA}. 
Therefore it becomes reasonable to consider
the polarizabilities as a suitable and sound framework to determine
the behavior of the symmetry energy with the parameters of the 
nuclear equation of state.
Developments with relativistic models also yield strong effects
with the isovector mesons, see for example references
\cite{GAITANOS,GRECOetal,IJMPD} among others. 
The density dependence of the neutron-proton symmetry energy
and the isospin dependence of the nuclear equation of state
are being extensively investigated  for several reasons
and several experimental tests are being done and prepared nowadays
mainly in intermediary and high energy heavy ion collisions
\cite{BAL-CMK-WB,SCALING-FRAG,ONO-DANIELEW-,RIA-GSI,IPHIC,BOTVINAetal,JYLIUetal,COLLECTFLOW4rho,EMITTEDPARTICLE,chin}.
The investigation of the possible effects  with
their particularities and the consequences for the 
observables is  extense involving new  experimental
facilities such as RIA and GSI \cite{RIA-GSI,IPHIC,GAMMA,BAOANLI} besides
many other works \cite{STEINERetal}.
Any definitive realistic investigation at really high densities 
(several times  the saturation density) should take into 
account 
baryonic structure with the internal (quark and gluon)

In the present work some aspects of the
symmetry energy terms are investigated
as provenient from the generalized polarizabilities
of nuclear matter for different ranges  of the  
density, n-p asymmetry and  momentum exchange, 
in the zero energy limit of the dynamical polarizabilities, within
general arguments and with
Skyrme forces at finite temperature. 
The case in which there is also 
non zero energy exchange corresponds 
to the analysis of the dynamical response function.
The density dependence of 
the equation of state is not
well known and it is reasonable to ask whether and how the 
n-p symmetry energy (and more generally other 
symmetry energies in the other channels of the 
nuclear interaction) 
depends on isospin  at different densities / very high n-p asymmetries.
The parabolic approximation, usually  appropriated for a restricted range of
densities (very) close to the saturation $\rho_0$ and small asymmetries,
may be modified for lower and/or higher densities.

The parameters of the forces which are used (SLyb and SKM) were fitted 
(i) from  results of asymmetric nuclear matter and 
neutron matter properties obtained from microscopic calculations 
\cite{CHABANAT} and (ii) from properties
of giant collective modes in $^{208}Pb$ 
 \cite{KRIVINE}. 
Other forces will be investigated elsewhere.
 Skyrme forces can be obtained from a reduction of the
nuclear density matrix \cite{BRINKVAUTHERIN-NEGELEVAUTHERIN} and their 
 basic structure is also
 present in non relativistic reductions of relativistic models
 for nuclear systems by passing to relativistic point coupling models or not
 \cite{SULAKSONOetal,BENDERetal}
 such that the necessary density dependence of each of the terms are
 expected to be stronger than considered in the earlier parametrizations
 \cite{FARINE,ONSIPP}.

This work is, in part,
an extension of  previous works 
 and it is 
organized as follows.
In the next section general aspects for the investigation of
symmetry energy within the approach of the general polarizabities 
are discussed including the  stability of nucleonic matter
with respect to external perturbations.
In section 3 an analysis of simultaneous dependence of the 
polarizabilities on the
neutron-proton asymmetry and on the density is proposed with a
differential equation that constrains these two behaviors of the 
symmetry energies.
In sections 4 and 5 the $q$-dependence (exchanged momentum between
the components of nuclear matter, eg. neutrons and protons)
of static generalized polarizabilities at finite temperatures 
 with  Skyrme forces is investigated 
in the limit of symmetric
nuclear matter. 
In the last section results are summarized.

\section{ Symmetry energy and nuclear matter polarizabities }

Basically, in this section, arguments from previous works are 
reproduced.
Consider that with the inclusion of an 
 external source of amplitude $\epsilon$, 
which separates nucleon densities 
with quantum numbers $(s,t)$ (where
$(0,1)$ is for spin up-spin down and
 $(0,1)$ for neutron-proton), 
the energy density of nuclear matter can be written as:
\be \ba{ll} \label{energ}
\displaystyle{ H = H_0 + \cA_{s,t}
\frac{( \rho_{(s,t)_1}-\rho_{(s,t)_2})^2}{\rho} +
\epsilon' \beta ,}
\ea
\ee
where $H_0$ does not depend on the density asymmetry
$(\rho_{(s,t)_1}-\rho_{(s,t)_2})^2$,
$\cA_{s,t}$  is the corresponding 
symmetry coefficient ($\cA_{1,0}$ the spin one, $\cA_{0,1}$ the neutron-proton
one) and the total density fluctuation is
$\beta = \delta \rho_{(s,t)_1}- \delta \rho_{(s,t)_2}$, 
for these two cases. 
For the spin-isospin external perturbation ($s,t = 1,1$)
the simultaneous fluctuations of the spin (up/down) and 
neutron/proton densities are to be considered just like
it is shown in expression (\ref{2}).
In the case of $(s=0,t=0)$, the scalar
channel, there is a change in the total nuclear density and 
$\cA_{0,0}$ is associated to a dipolar 
incompressibility \cite{NPA2000-2001}. 
For equal volumes the densities become the nucleon-numbers.

In the ground state, the variation of the energy 
with respect to the density fluctuation of a channel ($s,t$),
 $\delta \rho \equiv \beta$, 
yields the condition of minimum:
\be \ba{ll} \label{inducingfields}
\displaystyle{ 
\epsilon' + 2 \frac{\cA_{s,t}}{\rho} (\rho_m + \delta \rho ) \;\; = \;\;
\epsilon + 2 \frac{\cA_{s,t}}{\rho} \: \delta \rho \; = \;\; 0 ,
}
\ea
\ee
where $\rho_m = \rho_0^n - \rho^p_0 \neq 0$ is for
an n-p asymmetric matter (or correspondingly 
$\rho_m^s = \rho_0^{up} - \rho^{down}_0 \neq 0$ for
spin polarized matter)
and the total "inducing perturbation" for n-p asymmetric systems
is denoted by 
\be \label{redef} \ba{ll}
\displaystyle{ \epsilon = \epsilon' + 2 \cA_{s,t} \: \rho_m .}
\ea \ee
The ground state can be considered to have a 
(polarized) spin up-down asymmetric density given by 
 $\rho_m^s \neq 0$ simultaneously to
(or instead of) the n-p asymmetry (also denoted by $\rho_m$ above).
$\rho_m$ will be considered in most part of this paper .
The nuclear matter polarizabity in the channel $(s,t)$ can be 
written for $\epsilon'$ or for 
the total (inducing) perturbation, $\epsilon$,
respectively  as: 
\be \ba{ll} \label{Aqq}
\displaystyle{  
\Pi^{s,t}_a \equiv \frac{\beta}{\epsilon'}
= - \frac{\rho}{2 \cA_{s,t}^a}, \;\;\; (i)
\;\;\;\;\;\;\;
\Pi^{s,t} \equiv \frac{\beta}{\epsilon}
= - \frac{\rho}{2 \cA_{s,t}} \;\;\; (ii). }
\ea
\ee 
The stability condition for these expressions are different.
This will be discussed below.

The main development will be focused 
for the neutron-proton symmetry energy ($s,t = 0,1$)
although it is analogous for the other channels.
The neutron proton asymmetry used in the present 
work is defined by the neutron and proton densities
$\rho_n, \rho_p$ as:
\be \ba{ll} \label{b}
\displaystyle{ b = \frac{\rho_n}{\rho_p} - 1.}
\ea
\ee
An asymmetry coefficient
which is probably more familiar to the reader is given by:
\be \ba{ll} \label{alpha}
\displaystyle{
\alpha = \frac{(2\rho_{n}-\rho)}{\rho} . }
\ea 
\ee
They are related by: $b = 2\alpha / (1- \alpha )$.
The coefficient $b$ varies from $b=0$, 
in symmetric nuclear matter, up to $b \to \infty$, 
in neutron matter.
For the sake of generality 
the coefficient $\cA_{s,t}$ is considered 
to be a function of the density fluctuation $\beta$. 
The fluctuation  $\beta$ is  considered 
to depend on the n-p asymmetry $b$.
These parameters may be related to each other
and therefore it will be written that 
${\cA}_{s,t} = \cA_{s,t} (\beta)$.
Relations between $b$ and $\beta$ have been investigated 
by means of prescriptions.
Among those, one which leads to reasonable results is:
\be \label{relac} \ba{ll}
\displaystyle{
\beta = \delta \rho_n \left( \frac{2+b}{1+b} \right),}
\ea
\ee
Where $\delta \rho_n$ is the neutron density fluctuation. 
In the n-p symmetric limit $\beta = 2 \delta \rho_n$ and in another limit, 
in neutron matter, $\beta = \delta \rho_n$.
This ansatz (expression (\ref{relac})) 
is based on the assumption that
the density fluctuations are proportional to the respective 
density of neutrons and protons, i.e., 
$\delta \rho_n/ \beta = \rho_n / \rho$, being $\rho$ the total 
density.

The resulting expression for the  symmetry energy coefficient 
$\cA_{s,t}$ for the prescription above is given by
\cite{NPA2000-2001}:
\be \ba{ll} \label{Ab-solucao}
\displaystyle{ {\cal A}^{s,t} = 
{\cal A}^{s,t}_{sym} \frac{ 2 + 2 b}{2 + b} 
,}
\ea \ee
for a general $s,t$ (spin,isospin) channel of the effective
interaction.
In this expression  $\cA_{sym} = a_{\tau} \simeq 30 MeV$ is the 
s.e.c. of symmetric nuclear matter ($b=0$). 
The (generalized) n-p symmetry energy term
can be rewriten as: 
$$\cA^{0,1} = \cA^{0,1}_{sym}  (1 + \alpha),$$
which corresponds to a third 
order term in the binding energy, being
smaller than the quadratic term
because  $\alpha < 1$. 
In not very n-p asymmetric systems, those
with n-p asymmetry close to the stability line, $\alpha^3 << 1$.
For $b= 2$ ($\alpha = 0.5$, 
neutron density three times larger than the
proton density) it follows $\cA = 1.5 \cA_{sym}$. 
In the limit of neutron matter 
${\cal A} (b \to \infty) = 2 {\cal A}_{sym}$.
 For proton excess $b < 0$. 
Prescription (\ref{relac})
is therefore model-dependent and different choices yield 
other forms for the
the (asymmetric) static generalized ``screening functions''.
The  dynamical response functions are less sensitive to this
prescription.

So far it has been assumed that 
the stable density $\rho$ is independent of $b$ (or $\alpha$). 
Below it is envisaged a development to guide the 
simultaneous variations of these two
variables.
Several investigations of the role of the symmetry energy on 
observables in Radioactive Ions are being prepared for 
RIA and GSI. 
For this the density dependence of the symmetry energy is 
extremely relevant.
However for mass formulae of 
very asymmetric nuclei and for the equation of state
 at  densities different from $\rho_0$ 
the isospin dependence
of the symmetry energy may be different from the usual
one given by the {\it parabolic approximation}, expression (\ref{1}).
Furthermore, more elaborated 
pictures in relativistic mean field calculations, 
which considers isovector mesons,  $\vec\delta$,
yield a qualitative increase
of  the relevance of the neutron-proton asymmetry, 
with a larger difference 
of neutron and proton effective masses \cite{KUTSCHERA,LIU,GAITANOS,IJMPD}.
Experimental
bounds on the neutron and proton effective masses 
\cite{BAOANLI-mass}
may shed light on this.
The spin and spin-isospin symmetry energies can be
investigated analogously. 
For example, the behavior of
$\cA_{1,0}$ of the spin channel 
(which has already been written as $a_{\sigma}$ in the static
limit in the framework of the Landau's Fermi liquid theory), 
at variable densities was investigated in
 different works \cite{BJP-IJMPE,FSS2001,KAISER}.

\subsection{ Stability conditions}

In the usual case in which the density 
$\rho$ is not dependent neither on $b$ nor on $\beta$ 
there are two ways of writing
a solution for the polarizability $\Pi_{s,t}$ from expression 
(\ref{energ}).
They correspond to the different definitions 
of the external source shown before, respectively $\epsilon'$ $(i)$ and 
$\epsilon= \epsilon' + \rho_m$ $(ii)$.
They allow for defining polarizabilities given respectively by:
\be \label{solucao} \ba{ll}
\displaystyle{ 
\frac{\cA_{s,t}^a}{ \rho_0} = 
\frac{C^{s,t}}{(\Pi_{s,t}^a)^2} 
- \frac{1}{\Pi_{s,t}^a} \;\;\;\; (i), 
\;\;\;\;\;\;\;\;
 \Pi_{s,t} = - \frac{\rho_0 }{2 \cA_{s,t}} 
\;\;\;\; (ii), }
\ea
\ee
where 
$C^{(s,t)} = - \frac{\rho_0}{4(\cA_{s,t}^{sym})^a}$ is a constant, 
with the usual value of the symmetry energy coefficient. 
In the n-p channel: 
$(\cA_{0,1}^{sym})^a \simeq 30$ MeV (symmetric limit).
These two polarizabilities (\ref{solucao}) are equal
in the limit of symmetric nucleonic matter $\rho_m = 0$.
This derivation applies for any of the channels $(s,t)$.

Consider that the binding 
energy is to be minimized with respect to the 
density fluctuation $\beta$. 
From this an equilibrium 
condition for nuclear matter is obtained
with  $\delta^2 (E/A)/ \delta \beta ^2 > 0$, 
being different from other 
ones and complementary to them \cite{MULLER-SEROT}.
To be a stable minimum of the binding energy 
the coefficients, of both definitions of the polarizabilities, 
satisfy respectively:
\be \label{cond} \ba{ll}
\displaystyle{ \frac{\delta^2 E/A}{\delta \beta^2_{s,t}} = 
- \frac{2 C_{s,t}}{(\Pi_{s,t}^a)^2} > 0, 
\;\;\;\;(i), \;\;\;\;\;\;\;
 \frac{\delta^2 E/A}{\delta \beta^2_{s,t}} = 
2 \cA_{s,t} > 0 , \;\;\;\;(ii) , } 
\ea
\ee
where $\Pi_{s,t}^a$ and $\cA_{s,t}$ given in expressions
(\ref{solucao} (i)) and (\ref{solucao} (ii)) respectively.
The constant $C^{(s,t)}$ may be negative 
(stable symmetric nuclear matter 
\cite{BJP-IJMPE}) or positive,
$\cA_{s,t}$ and $\Pi_{s,t}$ also may be negative or positive.
While the second expression yields the more expected result,
i.e. the stability is directly shown by the signal of $\cA_{s,t}$ in 
each channel $(s,t)$, the first
expression has a more involved behavior due to the complicated
form of expression (\ref{solucao} (i)).
From the polarizability (\ref{solucao}-(i))
two conditions for real $\cA$ and stable
system follow: 
\be \ba{ll} \label{cond2}
\displaystyle{ 
(i) \; (1) \;\; \cA^a_{s,t} < (\cA^a_{s,t})_{sym}, \;\;\;\;\;
(i) \; (2) \;\; (\cA^a_{s,t})_{sym} \left( 2 - 
\frac{\cA^a_{s,t}}{(\cA^a_{s,t})_{sym}} \pm
2 \sqrt{1 - \frac{\cA^a_{s,t}}{(\cA^a_{s,t})_{sym }} } \right)  < 0.}
\ea
\ee
From condition (\ref{cond2} (i)-(1))
the neutron-proton asymmetry can only lower the value of 
the generalized coefficient $\cA_{s,t}$
to keep the system stable with the use of polarizability
(\ref{solucao} (i)).
It is worth emphasizing that the two conditions (\ref{cond}) with
the respective definitions for $\Pi_{s,t}$ should not be
mixed.
If the polarizability from expression (\ref{solucao}-(ii)) is considered
the condition  (\ref{cond}-(ii))
is to be applied, otherwise inconsistent results arise.
Expression (\ref{solucao}- (ii)) is the usual form.
However if one considers solution (\ref{solucao}-(i))
the condition  (\ref{cond}-(i))
is to be applied, otherwise inconsistent results arise.
In particular in the case of the polarizabity given by 
expression (\ref{solucao}-(i))
there are several possibilities for
 the stability of the symmetric and the corresponding asymmetric
matter depending on $\cA_{s,t} > 0 $ (or $\cA_{s,t} < 0$) 
and $\cA_{sym}^{s,t} > 0$
(or $\cA_{sym}^{s,t} < 0$) in each of the channel $(s,t)$. 
The microscopic {\it in medium} nucleon 
interactions, in an  exact calculation, would give the  correct one.
Expression (\ref{Ab-solucao}) for $\cA_{0,1} (b)$ 
was found with
the  solution (\ref{solucao} (ii)) for $\Pi_{s,t}$.

The stability conditions of a Fermi liquid in the leading order,
in each channel of the interaction, correspond to 
a particular case of the above expression (\ref{cond} - (ii)).
They are  given by the denominator of 
a particular limit of the response function $\Pi_{s,t}$
which can be written as:
\be \ba{ll} \label{ferliq}
a_{s,t} = N_0 ( 1+ J_0^{s,t}) > 0,
\ea \ee
 where $J_0^{s,t}$ 
stands for any of $F_0, F'_0, G_0, G'_0$ respectively for 
the scalar ($s=0,t=0$), 
isovector ($s=0,t=1$ with $a_{\tau}$), 
spin ($s=1,t=0$ with $a_{\sigma}$) and 
spin-isovector ($s=1,t=1$ with $a_{\sigma \tau}$) 
channels \cite{BACKMANBROWN,FERLIQ}. 
These expressions contain the leading terms 
of the more general calculation. 
Within a non relativistic formalism with Skyrme type interactions
they can be written in terms of Landau parameters 
\cite{BVA,GIAIetal}.
Other considerations can be associated 
in different formalisms 
\cite{AYCOCHO,FSS2001,KAISER,SPINISOSPINdelta}.
A complementary and more general discussion 
for particular models will be done in a 
forthcoming work.

\section{ Simultaneous dependence on isospin and density}

Next it will be assumed
 that there is an implicit and {\it a priori} 
unknwon dependence of the
saturation density on the n-p asymmetry without any 
supposition about the microscopic origin for this,
 $\rho_0 = \rho_0 (b)$. 
From the general and usual expression 
for the polarizability (\ref{Aqq}) (or (\ref{solucao} (ii))) a differential 
equation for the simultaneous isospin and density dependence of the
symmetry energy (coefficients) $\cA_{s,t}$ will be derived.
Although expression (\ref{Aqq}) was also derived without 
considering a dependence of $\rho$ on $b$ it will be considered 
that this simple form is more general.
 The derivative of the polarizability
 $\Pi_{s,t}$, expressions (\ref{Aqq}),
with respect to $b$ is given by: 
\be \label{eq1} \ba{ll}
\displaystyle{ \frac{\partial \beta}{\partial b} =  \beta 
\left\{ \left( \frac{1}{\rho} - \frac{1}{\cA_{s,t}} 
\frac{\partial \cA_{s,t}}{\partial \rho} \right) 
\frac{\partial \rho}{\partial b}
-\frac{1}{\cA_{s,t}} \frac{\partial \cA_{s,t}}{\partial b} \right\}
.}
\ea
\ee
The variation $\delta \beta/\delta b$ is given by
 expression (\ref{relac}), the prescription for the relation
between the fluctuations.
This equation has other three derivatives a priori unknown 
which have to be consistent with 
the equation of state:  the derivatives $\partial \cA/\partial \rho$, 
$\partial \cA/\partial b$ and $\partial \rho/ \partial b$.
This expression is therefore to be equated to that of 
prescription (\ref{relac})  or, more generally,:
\be \label{eq2} \ba{ll}
\displaystyle{ \frac{\delta \beta}{\delta b} \equiv - \beta 
f(b) 
.}
\ea
\ee
The resulting equation is:
\be \label{eq3} \ba{ll}
\displaystyle{
\left\{ \left( \frac{1}{\rho} - \frac{1}{\cA_{s,t}} 
\frac{\partial \cA_{s,t}}{\partial \rho} \right) 
\frac{\partial \rho}{\partial b}
-\frac{1}{\cA_{s,t}} \frac{\partial \cA_{s,t}}{\partial b} 
\right\} = - f(b)
.}
\ea
\ee
This is one of the most relevant results of this paper.
This equation constrains the simultaneous dependence of the
symmetry energy on the density and on the nucleon density
asymmetry
(through the generalized coefficient $\cA_{s,t}$,
in the channel $s,t$, which is not a constant anymore)
\footnote{ The parameter $b$ however 
 may be replaced by the equivalent one for
the spin up-spin down asymmetry in polarized nuclear matter.
The same form is obtained for spin-polarized nuclear matter, by
interchanging the neutron-proton variables (from $s,t = 0,1$) to
spin up- spin down ones (with $s,t = 1,0$). 
An equivalent  
 prescription has to be provided for the  
spin density fluctuations.}.

The following cases correspond to the derivation of section 2
(expressions (\ref{solucao})): 
$$ \frac{\partial \rho}{ \partial b} = 0 , \;\;\;\; \mbox{and / or}
 \;\;\;\;
\frac{\partial \cA}{ \partial \rho} = \frac{\cA}{\rho}.$$
These  correspond to the use of 
 prescription given by expression (\ref{relac}) which yields the
function: 
\be \ba{ll} \label{fb}
\displaystyle{f(b) =  \frac{1}{(1+b)(2+ b)}.}
\ea \ee
For this prescription, which yields expression (\ref{Ab-solucao}) 
for $\cA (b)$,
it has been assumed that $\rho$ was independent of $b$, therefore, in 
that case: 
\be \ba{ll} \label{abeta}
\displaystyle{ \frac{1}{\beta}\frac{\partial \beta}{\partial b}  
= - \frac{1}{\cA} \frac{\partial \cA}{\partial b}.
}
\ea
\ee
In this case the behavior of $\cA (\rho)$
can be the one typical of 
relativistic models with the increase of (any of the) 
symmetry energy coefficient
with the increase of the nuclear density, i.e., $\cA_{s,t} \propto \rho$
 and in part of 
microscopic approaches, for which its value usually tends to a constant,
\cite{KAISER,PAL,BCK,BALDOBOMBURG,WIRINGA,ENGVIK,REID-PARIS}.
However this is not the most general and interesting case because 
 condition (\ref{abeta}) holds
 when $\rho$ is independent on $b$.

A slightly more general parametrization can be investigated.
For example, Heiselberg and Hjorth-Jensen 
\cite{HEISELB-HJENSEN,BAOANLI4}
have used the following expression for the density dependence of the 
symmetry energy which nearly summarizes results 
obtained from relativist models: 
\be \ba{ll} \label{gamma1}
\displaystyle{ 
E_{sym} = E_{sym}(\rho_0)  
\left( \frac{\rho}{\rho_0} 
\right)^{\gamma} ,}
\ea \ee
 where
$\gamma$ is a constant and $\rho_0$ the saturation density.
A variational calculation 
favor values of the order of $\gamma \simeq 0.6$ whereas
an analysis of heavy ion collisions experiments at low energies 
$\gamma \simeq 2$ 
\cite{HEISELB-HJENSEN,BAOANLI4}. 
From the differential equation
(\ref{eq3})
it will be considered  parametrizations given by:
\be \label{eq5} \ba{ll}
\displaystyle{ \frac{\partial \cA}{\partial \rho} = 
\gamma \; \frac{ \cA}{ \rho}, }\\
\displaystyle{ \cA = \cA_{sym} 
\frac{\alpha_1 + \alpha_2 b}{ \alpha_3 + \alpha_4 b}
,}
\ea
\ee
where $\alpha_i$ (i=1,2,3,4) are constants. 
When the asymmetry coefficient in 
expression (\ref{eq6}) reaches the value 
$$b = - \frac{\alp_1}{\alp_2}$$
the symmetry energy coefficient $\cA_{0,1}$ changes the sign
making the system unstable according to the 
stability condition (\ref{cond} (ii)).
The resulting equation, from the general equation (\ref{eq3}), 
for the density as a function of $b$ with the above parametrizations
is:
\be \label{eq6} \ba{ll}
\displaystyle{ \frac{\partial \rho}{\partial b} = 
\frac{\rho}{\gamma} \left( -f(b) + 
\frac{\alpha_1 + \alpha_2 b}{ \alpha_3 + \alpha_4 b }
\right)
.}
\ea
\ee
For positive
$\alpha_3$ and $\alpha_4$ the general solution is given by:
\be \label{eq7} \ba{ll}
\displaystyle{ 
{\rho}(b) = (b + 2)^{(\frac {1}{\gamma})}\,(1 + b)^{( - \frac {1}{\gamma})}
\,({ \alpha_3} + { \alpha_4}\,b)^{(\frac {{\alpha_1}\,{\alpha_4} - {\alpha_2}
\,{\alpha_3}}{\gamma \,{\alpha_4}^{2}})}\,
e^{(\frac {{\alpha_2}\,b}{\gamma \,{\alpha_4}})}\,{ B}
.}
\ea
\ee
In the limit of symmetric nucleonic matter, for $b$ the neutron-proton
density asymmetry, $b = 0$, 
the constant $B$ can be fixed in terms of $\rho_0$.
In neutron matter 
$\rho \to 0$ or $\rho \to \infty$.
For this to be finite, $\rho(b \to \infty) \to 0$, one must have 
 $\alpha_2/\alpha_4 \leq 0$ and 
$\alp_1 \alp_4 - \alp_2 \alp_3 \leq 0$.
A particular  solution appears for $\alpha_3/\alpha_4=-b$ 
which yields $\rho = 0$.
When this occurs $\alp_1$ and $\alp_2$ from expression (\ref{eq7})
have different signs.

For the  usual form for the symmetry energy term 
in which $\cA = a_{\tau}$ is independent of $b$, 
the resulting density as a function of 
the asymmetry $b$ from the differential equation
(\ref{eq3}) is given by:
\be \label{eq8} \ba{ll}
\displaystyle{ \rho (b) = C_0 \left( \frac{2+ b}{1 + b} 
\right)^{\frac{1}{\gamma}} 
,}
\ea
\ee
where $C_0$ is fixed by a boundary condition, for example
 $\rho (b=0)$, with a fixed value of $\gamma$.
From  this limit:  $C_0 = \rho_0 2^{1/\gamma}$
whereas in neutron matter $\rho (b \to \infty) = 
C_0$. 
The ratio of the density in 
these two limits is given :
\be \label{eq9} \ba{ll}
\displaystyle{ \rho (b \to \infty) = \frac{\rho (b=0)}{
 2^{\frac{1}{\gamma}} }
.}
\ea
\ee
However,  the parameter $\gamma$ from the 
parametrization (\ref{gamma1}) may be (assumed to) 
depend on  the neutron-proton
asymmetry coefficient $b$ (or equivalently $\alpha$).
In this sense, a modification in the usual
 symmetry energy dependence  on the n-p
asymmetry can be expected to be equivalent
to different values for the  parameter $\gamma$,
  at different densities,
in different experimental situations. 

A different form for the above equation (\ref{eq3}) 
can be written by considering that:
$\frac{\partial \rho}{\partial b} \equiv g(b, \rho),$ and 
$\frac{\partial \cA}{\partial \rho} \equiv h(\rho, b) \neq \frac{\cA}{\rho}
$.
The following differential equation appear for $\cA(b,\rho)$
with these functions:
\be \label{eq4} \ba{ll}
\displaystyle{
\left\{ \left( \frac{1}{\rho} - \frac{1}{\cA_{s,t}} 
h(\rho, b) \right) g(b, \rho)
-\frac{1}{\cA_{s,t}} \frac{\partial \cA_{s,t}}{\partial b} \right\} 
= - f(b)
.}
\ea
\ee
Considering the particular prescription (\ref{abeta}) 
it is  obtained the following expression:
\be \label{eq-elimin} \ba{ll}
\displaystyle{
\left( \frac{1}{\rho} - \frac{1}{\cA_{s,t}} 
h(\rho, b) \right) g(b, \rho)
= - 2 f(b)
.}
\ea
\ee
 
These expressions can be considered for any channel $(s,t)$.
They generate one differential
equation for each channel of the nuclear effective interaction with
$\cA_{s,t}$,
and therefore the final $\rho$ dependence on $b$ is to be the same
for each of these equations, 
for $b$ representing the same asymmetry 
(neutron-proton, spin-up-spin-down).
For this, the choices for $\cA_{s,t}$ and 
$\beta (b)$ should be associated, otherwise there will appear
different $\rho (b)$.

\section{ Generalized "Screening functions" with Skyrme forces }

In this section the analysis done previously
\cite{BVA,FLB99,NPA2000-2001,BJP-IJMPE} is extended with
the static limit of the
expression for the dynamical polarizability 
of a non relativistic hot asymmetric nuclear 
matter  with Skyrme effective
interactions,
$lim_{\omega \to 0} \Pi_{s,t} (\omega, q)$.
These polarizabilities were obtained by 
the calculation of the response function 
of hot asymmetric nuclear matter in terms of three densities: 
neutron and proton densities ($\rho_i$), momentum
density ($\tau_i$) and kinetic energy ({\bf j}$_i$) densities
from the time dependent Hartree Fock approximation with
Skyrme forces \cite{BVA,NPA2000-2001}.
These densities appear in reductions
from relativistic models in which the scalar density is written
in terms of them \cite{SULAKSONOetal,BENDERetal}.
The time dependent approach introduces CP violating terms
proportional to {\bf j} which 
 are larger in asymmetric nuclear matter.
Four asymmetry coefficients  are defined,
$a, b, c$ and $d$ for the effective masses and densities, and they
are given by:
\be \label{5c} \ba{ll}
\displaystyle{
a = \frac{m^*_{p}}{m^*_{n}} - 1 , \;\;\;\;\;
b = \frac{\rho_{0n}}{\rho_{0p}}  -1, \;\;\;\;\; 
c = \frac{1+b }{2+b }, }\;\;\;\;\;
\displaystyle{ d = \frac{1 }{1+ (1+b)^{\frac{2}{3} } } ,}
\ea
\ee
where $m_i^*$ are the neutron and proton effective masses.
Small  approximations were done:
 (i) to equate the asymmetry coefficient defined for the
momentum density to the density asymmetry coefficient
(ii)
to choose a particular  prescription for the fluctuations 
of the asymmetry density - expression (\ref{relac}).
The second approximation is in fact a choice with dynamical content
 and it deserves more attention.

At the Hartree Fock level, 
the symmetry energy coefficient,
$a_{\tau} = \partial^2 (E_0/A) \partial \alpha^2$, 
 can be written from  the expansion:
\be \ba{ll} \label{usualdef}
\displaystyle{ \frac{E_0}{A} = H_0 + \alpha 
\left. \frac{\partial (E_0/A)}{\partial \alpha} \right|_{\alpha = 0} +
\frac{\alpha^2}{2}
\left. \frac{\partial^2 (E_0/A)}{\partial \alpha^2} \right|_{\alpha = 0} +...,
}
\ea \ee
where higher order terms are not written.
There may appear (small) higher order terms. 
By calculating the general polarizability, within the linear response
approach, a whole class of ring diagrams contribute beyond the 
Hartree-Fock \cite{NEOR}. 
Therefore more complete symmetry energy terms can be obtained.

For the calculation of the response function 
an external source is introduced in the Hartree Fock
time-dependent equation which induces small amplitude density fluctuations.
The general form of the source is the plane wave one, it is given by:
\be \ba{ll}
\displaystyle{ V_{ext} = - \epsilon \; \hat{O}_{s,t}\;
 {\cal D} \;
e^{ - i (\omega t - \bq.\br)},
}
\ea
\ee
 with an amplitude $\epsilon$ (usually a small parameter),
 an associated  dipole moment ${\cal D}$
(equal to the unit from here on) and 
the operator $\hat{O}$ acts on the nucleon states. In particular, 
for the isovector interaction the third component of the
 isospin (Pauli) matrices is considered yielding
neutron-proton density fluctuations. 
With the above external source, the Hartree Fock equation 
for nuclear matter is written as:
\be \label{HFeq}
\displaystyle{ \partial_t \rho_i = - i 
\left[ W_i + V^{ext}_i , \rho_i \right]  },
\ee
where $W_i$ is  the Hartree-Fock energy of  protons or neutrons.
 The induced density fluctuations $\delta \rho$ are 
 to have the same spatial and temporal plane waves behavior 
 of the external source.

The resulting  expression 
is more appropriatedly written in terms of
 generalized Lindhard functions whose real parts, at zero 
temperature $F_{2i}$, were defined as \cite{BVA,FLB99}:
\be \ba{ll} \label{16}
\displaystyle{ \Re e \Pi_{2N}^{i}(\omega, {\bf q})  \equiv
\frac{g M^*}{2 \pi^2}
\Re e 
\int \hbox{d}^3 k  \frac{ 
 f_{q} ({\bf k} + {\bf q})- f_{q}( {\bf k})  }{
\omega +i \eta-  \epsilon_p' ({\bf k}) + \epsilon_p'({\bf k}+ {\bf q}) 
 }
 \left( {\bf k} .({\bf k +q }) \right)^N =
 \frac{g M^*}{2 \pi^2}
\int d f_i (k) \Re e \; F_{2i}
.}
\ea
\ee
In these expressions $f_i(k)$ are the
fermion occupation numbers for neutrons ($i=n$) and 
protons ($i=p$) which will be considered only for the
zero temperature limit (when $d f_i (k) \to - \delta (k- k_F)$), 
$g$ is the degeneracy factor
for spin and isospin, 
$M^*$ is the effective mass in symmetric nuclear matter. 
In the limit of zero energy exchange ($\omega \to 0$)
the Lindhard functions yield
the  (q-dependent) proton and neutron
 densities, momentum and kinetic energy
densities are given by:
\be \ba{ll} \label{densities}
\displaystyle{ N_q = \frac{\gamma M^*}{2 \pi^2}
\int d f_i (k) \; \Re e F_{0} (\omega \to 0), }\\
\displaystyle{ \rho_q = \frac{\gamma M^*}{2 \pi^2}
\int d f_i (k) \; \Re e F_{2} (\omega \to 0), }\\
\displaystyle{ M_q = \frac{\gamma M^*}{2 \pi^2}
\int d f_i (k) \; \Re e F_{4} (\omega \to 0).}
\ea
\ee

In  the symmetric nuclear matter
 the  momentum-dependent polarizability (\ref{Aqq})
in the channel $s,t$ for the Skyrme effective force 
parametrization is written as:
\be \label{ASYM} \ba{ll}
\displaystyle{A_{s,t} (q) = 
\frac{\rho^q}{ N^q}
\left\{ 1 + 2\overline{V_0^{s,t}} N^q
+6 V_1^{s,t} M^* {\rho^q} + 
 (V_1^{s,t})^2 (M^*)^2 \left( 9  {\rho_q}^2
- 4  M_q  N_q \right)  \right\}
, }
\ea
\ee
Where $\overline{V_0} (q^2)$ and $V_1$ are functions 
of the Skyrme forces parameters for each of the $(s,t)$ channel
shown in the Appendix.
The nuclear matter incompressibility modulus is related to
$A_{0,0} (q^2=0)$ in the Appendix.
The q-dependent densities $N^q, \rho^q, M^q$ are the total densities
from expressions (\ref{densities}).
 The term proportional to $V_1^2$ can be re-written in 
a homogeneous nuclear matter at zero temperature,
 as $\rho \tau - \vec{j}^2$ which is to be zero in the 
 Galilean invariant (homogeneous and static) limit
 \cite{BENDERetal}. 
This invariance is  broken in these 
 cases and it is amplified
in asymmetric nuclear matter.
$q$ in the neutron-proton channel  is the exchanged momentum between 
the neutron and proton components, and similarly, in the 
spin channel, the corresponding 
exchanged momentum for spin up and down nucleons.
The stability condition for this expression is given by
(\ref{cond} - (ii)). 
The finite temperature calculation of the densities
lead to finite temperature symmetry energy coefficients.
In the zero frequence limit the 
imaginary part of the response function
 disappears.

In the limit of low momenta, $q << 2 k_F$, the 
$w=0$ limit of the Lindhard
functions are simplified, as shown in the Appendix.
The polarizabilities of symmetric nuclear 
can be approximatedly written in the following form:
\be \ba{ll} \label{Aq1}
\displaystyle{ {\cA}_{s,t} =  
 \cA_{s,t}(T,\rho) + 
A^{(1)}_{s,t}(T,\rho) q + 
A^{(2)}_{s,t}(T,\rho) q^2 ,
}
\ea
\ee
where ${A}_{s,t}(T,\rho)$ is the usual symmetry 
energy coefficient in the channel $(s,t)$ \cite{BVA,NPA2000-2001} 
and ${A}^{(i)}_{s,t}(T,\rho)$ 
are functions of the
Skyrme force parameters 
(combined in the functions $V_0^{s,t}$ and $V_1^{s,t}$), $\rho$ and T.

\subsection{Results for Skyrme interactions in the 4 channels}

In this section the generalized
polarizabilities are investigated numerically
for the Skyrme forces SKM and SLy(b) for
 the four channels of the particle-hole interaction
 as functions of the 
exchanged momentum at the 
normal density $\rho_0$.
For this, the chemical potential 
was adjusted to mantain a constant stable nuclear density
$\rho (T) = \rho_0$. 
As a consequence the results are not very strongly
dependent on $T$.
In Figure 1 the neutron-proton polarizability 
is shown as a function of the (exchanged) momentum for temperatures 
$T = 0, 4$ and $7$ MeV for Skyrme force SLyb and at $T=0$ MeV for
the force SKM.
Both forces produce widely accepted values
for the symmetry energy coefficient, 
 $\cA_{0,1} (q=0, T=0) \simeq 32$ MeV.
There is a  general behavior (for both forces at any
of the temperatures)  of decreasing $\cA_{0,1}$ with 
increasing exchanged momentum up to $q \simeq 500$ MeV.
This corresponds to nearly
twice the nucleon momentum at the Fermi surface. 
However $\cA_{0,1}$ does not reach negative values.
The behavior of decreasing values of $\cA^{0,1}$
for increasing momenta
 is  in agreement with other
analysis for the momentum dependence of the
symmetry energy \cite{BAO-DAS-GUPTA-GALE}.
The  symmetry 
energy coefficient has, according to 
expressions (\ref{ASYM},\ref{Aq1}), a linear/quadratic 
behavior for very small exchanged momenta $q$. 
This
is followed by an
abrupt change of behavior at $q \simeq 500$ MeV.
For higher $q$ the generalized s.e.c. ($\cA_{0,1}$) increases 
nearly linearly 
at different temperatures.
For the s.e.c. $\cA_{0,1}$ to become negative the 
Skyrme parameters $t_0$ and $t_3$ should  
result in larger  values of $\overline{V}_0$ 
than those of SLyb (this variable
is still smaller for the force SKM) 
and/or different values for $t_1, t_2$.

In Figure 2 the spin-isospin generalized polarizability 
dependence on exchanged momentum between 
neutrons and protons with spin up and down 
is investigated for the
same cases of Figure 1. 
There is again a quite defined change of behavior at 
$q \simeq 500$ MeV.
The generalized spin-isospin s.e.c. remains
nearly constant with increasing $q$ up to $q \simeq 500$MeV. 
The force SKM yields smoother variations than SLyb
like  in the n-p channel. 
Above the saturation density, $\cA_{1,1}$ decreases 
for most forces eventually reaching a 
negative value  \cite{BJP-IJMPE} .
Finite temperature effects are larger for
higher values of $q$.

In Figure 3 the spin generalized polarizability, $\cA_{1,0}$,
 is shown for the same cases of the
previous figures. The turning
point present in  the isospin-dependent channels,
investigated in figures 1 and 2, 
is the same ($q \simeq 500$ MeV).
This is due to the form of the Lindhard functions.
However the behavior is completely different
for each of the forces that already have very different 
predictions of $\cA_{1,0} (q=0,T=0)$. 
SKM yields a nearly constant behavior followed by a strong
 increase of $\cA_{1,0} (q)$ for very large $q$ 
whereas SLyb decreases to a local minimum at 
$q = q_c \simeq 500$ MeV. 
The behavior resulted by the use of SKM force shows qualitative
 agreement with
the results by Kaiser within Chiral Perturbation Theory \cite{KAISER}.
For the  force  SLyb the spin  symmetry energy coefficient 
may decrease 
still more for large values of $q$,
at zero temperature, eventually it may become negative at larger
 densities.
The instability associated to $\cA_{1,0} < 0$ is the one
 towards a ferromagnetic alignment which has been found
in several works with several Skyrme forces and relativistic models
at higher densities 
\cite{VIDAURREetal84,BJP-IJMPE,ISAYEV,KUTSCHERAW1,BERNARDOSetal95,MARCOSetal91}.
However this transition is  
absent in several calculations.
The most well known calculations in which the ferromagnetic
alignment is not found are those ones based on 
NN interactions with different methods \cite{FSS2001,VB02,VPR,KAISER}.
However there are 
 particular Skyrme forces which do not provide this ferromagnetic 
phase for nuclear matter: 
those parametrizations
with the inclusion of NN tensor Skyrme-type force by Liu {\it et al}
 \cite{LIUetal91} or 
using SLyb
at low momentum as seen in figure 3, for higher densities 
and n-p asymmetries - seen in the second of the references
 \cite{BJP-IJMPE}.
The effect of the momentum dependence, however, is the
decrease of $\cA_{1,0} (q)$.
Whereas the functional density formalism with
Skyrme forces and the relativistic (mean-field) models with 
nucleon-mesons couplings are effective 
models for the
nuclear many body problem the NN based calculations are subject
to approximative methods which may not capture all the 
relevant degrees of freedom appropriatedly in each part of the nuclear 
phase diagram. 
At finite temperatures $\cA_{1,0} (q, T)$ does not vary significantly.

In the Figure 4 the scalar polarizability, $\cA_{0,0}$ 
as defined in expression (\ref{ASYM}) is plotted. 
It shows a continuous increase with momentum without the
turning point  at $q \simeq 500$MeV.
This parameter, a dipolar incompressibility, 
is proportional to the nuclear matter incompressibility, 
like it is shown in the Appendix.

In all examples shown above, the increase of temperature is more relevant for 
larger $q$ and the increase of the nuclear temperature 
always yields larger $\cA_{s,t}$. 
Usually it is not 
expected a large variation of the static symmetry energy 
with the
temperature from microscopic calculations in finite nuclei
 \cite{DEAN,DONATI}.  
Modifications of the chemical potential at high temperatures can 
lead to stronger dependences on $T$.

To understand the behavior of the generalized polarizabilities
the total densities ($N^q, \rho^q, M^q$), as defined above, 
are shown in figure 5 as functions
of $q$ for the parameters of force SLyb. 
They are obtained from the zero-frequence of the
generalized Lindhard functions 
and they  generate the behavior of expressions (\ref{Aq1})
seen in figures 1 to 4.
Whereas $\Pi_0$ and $\Pi_2$ present a smooth
behavior towards to zero with the increase of $q$, 
the momentum density, $|\vec{j}| \propto M$, has a dramatic
changement at $q \simeq 2 k_F$. 
This 
prevents the polarizabilities $\cA_{s,t}$
to become negative
for the forces investigated in this work, in particular the
neutron-proton 
$\cA_{0,1}$ and spin $\cA_{1,0}$ ones.

\subsection{Other considerations}

 From  the stability analysis of section 2,
 the results shown in figures 1 to 4,
mainly for the force SLyb,  
suggest that nuclear matter is close to
undergo  phase transitions around
$q \simeq 500$ MeV, i.e., when the
exchanged momentum $q$  is nearly twice the 
momentum at the Fermi surface, $k_F$.
The q-dependence of the Lindhard function yield  $M_i(q)$, 
as well as $N_i(q), \rho_i(q)$ for $\omega =0$,
which
prevents nuclear matter to undergo phase transitions.

This analysis was done for zero energy 
with $\cA_{s,t}$ only as a function of 
exchanged momentum.
The frequence dependence of the polarizabilities was analysed
associatedly to the exchanged momentum for the
 dipolar 
collective motions  where
zero-sound like excitations  were found 
\cite{BVA,NPA2000-2001}. 
In nearly symmetric nuclear matter
they disappear at temperatures of the order of $T \simeq 7$ MeV
\cite{GR,BRAGHINVAUTHERIN} (and higher temperatures for 
non zero asymmetries).
Their disappearance may occur with
 the  liquid-gas phase transition \cite{POCHO,CHOMAZ-SIM}.
The increase of the giant dipole isovector resonance width 
stops so that the corresponding energy is probably being used for
changing the phase of the system

These results can also be expected to yield consequences for the
Supernovae mechanism and proto-neutron or "neutron" stars with
their dynamical behavior involving
energy and momentum dependence of $\cA_{s,t}$. 
The symmetry energies contribute, among other ways, by means of
$\cA_{0,0}$, $\cA_{1,0}$ (due to the coupling to neutrinos) and 
$\cA_{0,1}$, $\cA_{1,1}$ for the different neutron-proton densities and 
the other related effects
\cite{BETHE,REDDY}.
The neutronization of a proto-neutron star in the
quasi-static phase of the supernova can be partially 
suppressed due to the eventual increase of the symmetry energy 
coefficient although the momentum dependence shown in Figure 1 
presents the opposite trend of decreasing $\cA_{0,1}$ up to
$q \simeq 2 k_F$. 
This second (dynamical) effect seemingly would
facilitate the neutronization and it should compete with the former. 
On the other hand the spin symmetry energy is strongly dependent
on the used Skyrme interaction.
Although the (continuous) increase of $\cA_{1,0}$ seems to be rather
in agreement with other works \cite{FSS2001,KAISER}
new developments are needed and they include
the need of new parametrization of effective forces
 focusing on spin-dependent
observables from nuclei and nuclear matter.

\section{ Summary}

In this paper the  nuclear matter symmetry energy 
terms were 
investigated as generalized polarizabilities.
Stability conditions with respect to neutron-proton 
fluctuations were derived
being complementary to others usually investigated 
\cite{FERLIQ,MULLER-SEROT}.
A differential equation for the simultaneous dependence 
of the generalized symmetry energy coefficients on the
neutron-proton asymmetry and on the total nuclear density
was proposed with equation  (\ref{eq3}).
For this, no considerations about the microscopic reasons for 
the resulting  stability density
 with a given n-p asymmetry  were raised.
The stability density is, in this case, to be a 
function of the n-p asymmetry
 as it should be in a general formulation. 
Some solutions for this equation were given.
This procedure is interesting for finite nuclei as well.
These results may be of interest for the 
investigations of the role of the symmetry energy on 
observables in Radioactive Ions which are 
 being prepared and done mainly for the 
RIA and GSI  machines. 
At different densities the isospin dependence
of the symmetry energy may be different from the usual one.
Finally, within the framework of the 
linear response of non relativistic nuclear matter with Skyrme forces, 
 the $q$-dependence (exchanged momentum between
the components of nuclear matter, eg. neutrons and protons) 
of the coefficients $\cA_{s,t}$ was investigated.
For low momenta the n-p symmetry energy 
decreases (linearly and quadratically)
 until $q_c \simeq 500$ MeV in agreement
with earlier investigations of the symmetry energy potential
\cite{BAO-DAS-GUPTA-GALE}. 
In this range of momentum transfer $q$  phase transition(s)
 may take place 
if other conditions are present, such as different (higher/lower)
nuclear densities.
This indication can be
seen in the other symmetry energies, the spin-dependent ones,
which however depend strongly on the particular Skyrme
effective interaction.
The results in the spin-dependent channels show no defined
sign of such ferromagnetic phase  for the SLyb force at the
saturation density. 
However the decrease of $\cA_{1,0} (q)$ with increasing
transferred momentum may favor such phase transition 
in different conditions of densities and n-p asymmetries.
The scalar coefficient $\cA_{0,0} (q)$, the
dipolar incompressibility,  has continously larger values
with the increase of exchanged momentum.

\vskip 0.5 cm
\setcounter{equation}{0}
\renewcommand{\theequation}{A.\arabic{equation}}
\noindent {\Large {\bf Appendix: Skyrme force parametrization, 
functions $V_i$, relation between $K_{\infty}$ and $\cA_{0,0}(q=0)$ 
 }}
\vskip 0.2cm

In this appendix we exhibit the functions $V_i$ (for expression
(\ref{ASYM}) with parametrization of 
Skyrme forces SKM, SLyb and others \cite{ONSIPP} given by:
\begin{equation} \label{skyrme} \begin{array}{ll}
v_{12} = &{\displaystyle
 t_0 ( 1 + x_0 P_{\sigma} ) \delta ({\bf r_1} - {\bf r_2}) +
\frac{t_1}{2} \left(1+ x_1 P_{\sigma} \right) \left[ \delta({\bf r_1} -
{\bf r_2}) k^2 + k'^2 \delta (
{\bf r_1} - {\bf r_2}) \right]  + }\\
&{\displaystyle
t_2 \left( 1+ x_2 P_{\sigma} \right) {\bf k'}. \delta ({\bf r_1} - {\bf r_2})
{\bf k}  + \frac{t_3}{6} ( 1 + x_3 P_{\sigma} ) (a_1 
(\rho_1 + \rho_2)^{\gamma} 
+ a_2 \rho^{\alpha}) \delta
({\bf r_1} - {\bf r_2}) ,}
\end{array}
\end{equation}
where $P_{\sigma}$ is the spin
exchange operator. 
The parameters for the forces SLyb and SKM are given respectively 
in references \cite{CHABANAT,KRIVINE}.

From the linear response calculation for a time dependent
Hartree Fock frame, in the lines
discussed in \cite{BVA,FLB99}, we can write
the corresponding
functions $V^0_{s,t}$ and $V^1_{s,t}$ 
in each channel for the more general calculation
in asymmetric nuclear matter:
\begin{equation} \label{V0V1}\begin{array}{ll}
\displaystyle{
\overline{V_0^{0,1}}= \left(
- \frac{t_0}{2} \left(x_0+ \frac{1}{2} \right) - \frac{t_3}{12}
\left[ a_2 \left( x_3 + \frac{1}{2} \right) 
+ a_1 \left(1 + \frac{x_3}{2}\right) -\frac{1}{4}(1-x_3)(\alpha+2)
(\alpha+1) \right]
 \rho^{\alpha}  + \right. }\\
\displaystyle{ \left.
- \frac{q^2}{16} ( 3 t_1
( 1+2 x_1) + t_2 ( 1+ 2 x_2) )
 \right) (1 + bc) + V_2^{0,1}, }\\
\displaystyle{ V_1^{0,1}
= \frac{1}{16} ( t_2( 1+ 2 x_2) -  t_1( 1 + 2 x_1) ) ,} \\
\displaystyle{ V_2^{0,1} = t_3\left[ 
a_2 (\frac{1}{2}+x_3)\alpha \rho^{\alpha-1} (c\rho_n + (c-1)\rho_p ) +
\right. } \\
 \displaystyle{ \left.
+ a_1 \left( (1+\frac{x_3}{2})\alpha \rho^{\alpha-1} (c\rho_n + \rho_p(c-1) )
 + 2 (1- x_3) (\alpha+2) (\alpha+1) (c\rho_n^{\alpha} + 
\rho_p^{\alpha} (c-1) )\frac{1}{16}
\right)
 \right] \frac{1}{12}   ,} \\
 \displaystyle{
\overline{V_0^{0,0}}
= \left( 3 \frac{t_0}{4} + (\alpha+1)(\alpha+2) t_3
\rho^{\alpha} \left[ a_1 (1+\frac{x_3}{2})
\left(\frac{1+ b}{2+b}\right)^2\frac{1}{16} + a_2 (1+ \frac{x_3}{2})
\frac{1}{12} \right]
  + \right. }\\
\displaystyle{ \left.
+ q^2 ( 9 \frac{t_1}{32} - (5+4 x_2) \frac{t_2}{32} ) 
\right)(1+bc) + V_2^{0,0}
},\\
 \displaystyle{ V_1^{0,0}
=  3\frac{t_1}{16} + (5 +4 x_2) \frac{t_2}{16}
}, \\
 \displaystyle{ V_2^{0,0} = \frac{t_3}{12} \left\{
(x_3+.5)(c \rho_n + (c-1)\rho_p
 \rho^{(\alpha-1 )}) +  \right. }\\
\displaystyle{ \left. +
a_1 \alpha (1-x_3)\left( 
\frac{(2\rho)^{\alpha}}{(2+b)^{\alpha+2}} + 
2\frac{((1+b)^2\rho)^{\alpha}}{(2+b)^{\alpha+2}}  \right)\frac{1}{2}
-a_2 \frac{( 1 + (1+b)^2\rho^{\alpha} )}{(2+b)^2} \right\}
}\\
 \displaystyle{
 \overline{V_0^{1,0}} = \left( -.5 t_0 (x_0+.5) - \frac{t_3}{12}
\rho^{\alpha}(.5+x_3) 
   - \frac{q^2}{8} 
( t_2*(x_2+.5) + 3 t_1 (.5+x_1) ) \right. }\\
\displaystyle{ \left.
 + \frac{a_1}{12} t_3 x_3 \rho^{\alpha} (2 +\alpha)  
 + \frac{a_2}{24}  t_3\rho^{\alpha} (2x_3-1)
 \right) ( 1 + b.c) + V_2^{1,0}
,} \\
 \displaystyle{
V^1_{1,0} =  \frac{1}{8} \left(t_2( x_2+ .5) - t_1(x_1+.5) \right) 
}\\
\displaystyle{ V_2^{1,0} = 
\frac{t_3}{12} (.5+x_3)\rho_n \rho_p^{(\alpha-1)} \alpha ).c + 
  t_3 (.5+x3)\rho_p \rho^{(\alpha-1)}\frac{\alpha}{12} (c-1) ,
}\\
 \displaystyle{ \overline{V_0^{1,1}} (q^2) 
= \left(-\frac{t_0}{4} - \frac{t_3}{24} \rho^{\alpha}
 - \frac{a_1}{48} t_3((2 \rho_n)^{\alpha}+ (2 \rho_p)^{\alpha})
- \frac{a_2}{24} t_3 \rho^{\alpha} + \right.  }\\
\displaystyle{ \left.
+ q^2 (-3 \frac{t_1}{32} - \frac{t_2}{32}) )(1 + b . c) \right)  + V_2 ,}\\
\displaystyle{ V_1^{1,1} = -\frac{t_1}{16} + \frac{t_2}{16}, }\\
\displaystyle{ V_2^{1,1} = \frac{\alpha}{24} t_3 \rho^{(\alpha-1)} 
(\rho_n  . c - \rho_p (1- c )  )
  - a_1 t_3 (2 + \alpha) ( (2 \rho_n)^{\alpha} c + 
  (2 \rho_p)^{\alpha} (c -1) )\frac{1}{12} + }\\
\displaystyle{ 
 + a_2 t_3 \alpha \rho^{(\alpha-1)} (-\rho_n \frac{c}{2}
    -(c -1) \frac{\rho_p}{2} )
(\rho_n .c  - \rho_p (1 -c)) \frac{1}{24}  ,}
\end{array}
\end{equation}
where $\rho_{n}$,  $\rho_{p}$  and $\rho$ are the 
proton, neutron  and total 
densities of  asymmetric nuclear
matter, $a, b, c$ are the asymmetry coefficients
defined in section 4.

For the longwavelength limit of 
$\cA_{0,0} (b=0, \rho, q^2 = 0)$  
the following relation  is obtained in terms of the 
incompressililiyt modulus \cite{NPA2000-2001}:
\be \label{K-A} \ba{ll}
\displaystyle{ \frac{1}{9} K_{\infty} = \cA_{0,0}  -  \frac{4}{5}T_F 
+ 2V_1 k_F^2 \rho_0 - \frac{3}{4} t_3 \rho_0^{\alpha +1}
.}
\ea \ee
They have different relevant terms such as the one proportional
to the nucleon kinetic energy at the Fermi surface, $T_F$, and 
a term from the density dependence of the Skyrme forces
proportional to $t_3$. 
This can be seen, in general, by remembering that
the calculation of $\cA_{0,0} (\rho,q=0)$ was done with the quadratic 
form for the binding energy in the presence of an external 
perturbation 
which induces density fluctuations  of expression (\ref{energ}).
It is rewritten below:
\be \ba{ll} \label{energ10}
\displaystyle{ H = H_0 + \cA_{0,0} (\rho)
\frac{( \delta \rho_{0,0} )^2}{\rho} +
\epsilon' \delta \rho.}
\ea
\ee
Terms containing $(\delta \rho)^n$, for $n \neq 2$, in $H$
were neglected and would correspond to the terms
which yield the usual $K_{\infty}$.
This more general parametrization will be considered in 
a forthcoming work.

The general structure of the zero frequence (real)
generalized Lindhard functions $\Pi_{2N}$ at zero temperature
 can be written as:
\be \ba{ll} \label{lindfunc}
\displaystyle{ \Pi_0 (T=0) = \frac{ M^* k_F}{\pi^2}
\left( -1 +  \frac{a}{q} (1-  \frac{q^2}{2 k_F^2}) 
Ln \left| \frac{q - 2 k_F}{q + 2 k_F}\right|
\right)
,}\\
\displaystyle{  \Pi_2 (T=0) = \frac{ M^* k_F^3}{2 \pi^2}
\left( -3 + 3 q^2 b^2 +  \frac{b}{q} ( 1 - 3 c q^2)
(1-  \frac{q^2}{2 k_F^2}) 
Ln \left| \frac{q - 2 k_F}{q + 2 k_F}\right|
\right)
, }\\
\displaystyle{  \Pi_4 (T=0) =  \frac{ M^* k_F^5}{ \pi^2}
\left( a_4 + b_4 q + c_4 q^2 + d_4 q^3+ e_4 q^4+  \frac{1}{3 q} 
( 1 - \frac{q^6}{(2 k_F)^6})
Ln \left| \frac{q - 2 k_F}{q + 2 k_F}\right|
\right),
}
\ea \ee
where $a_i,b_i,c_i,d_i,e_i$  dependend on $k_F$ and  on
$M^*$.

\vspace{2cm}

\noindent {\Large {\bf  Figure captions}}

\vskip 0.5cm

{\bf Figure 1 } Neutron-proton symmetry energy coefficient 
 $A_{0,1}= \rho/(2 \Pi_R^{0,1})$ of symmetric nuclear matter
 as a function of the momentum transfer between 
neutrons and protons, 
$q$(MeV), for interactions SLyb 
for $T=0, 4, 7$ MeV and SKM (T=0).

{\bf Figure 2} Spin symmetry energy coefficient 
 $A_{1,0}= \rho/(2 \Pi_R^{1,0})$ of symmetric nuclear matter
 as a function of the 
momentum transfer between 
neutrons and protons, $q$ (MeV), for interaction SLyb 
for $T=0, 4, 7$ MeV and SKM (T=0).

{\bf Figure 3 } Spin-isospin symmetry energy coefficient 
 $A_{1,1}= \rho/(2 \Pi_R^{1,1})$ of symmetric nuclear matter
 as a function of the 
momentum transfer between 
neutrons and protons, $q$ (MeV), for interaction SLyb 
for $T=0, 4, 7$ MeV and SKM (T=0).

{\bf Figure 4} Scalar symmetry energy coefficient 
 $A_{0,0}= \rho/(2 \Pi_R^{0,0})$ of symmetric nuclear matter
 as a function of the momentum transfer  between 
neutrons and protons,
$q$ (MeV), for interaction SLyb 
for $T=0, 4, 7$ MeV and SKM (T=0).

{\bf Figure 5} The densities $N, \rho, M$ as functions of 
the transfered momentum between neutrons and protons
for the force SLyb. 
They are nearly independent of the 
force unless for the values of $m^*$ and $k_F$.

\vspace{2cm}

\begin{figure}[htb]
\epsfig{figure=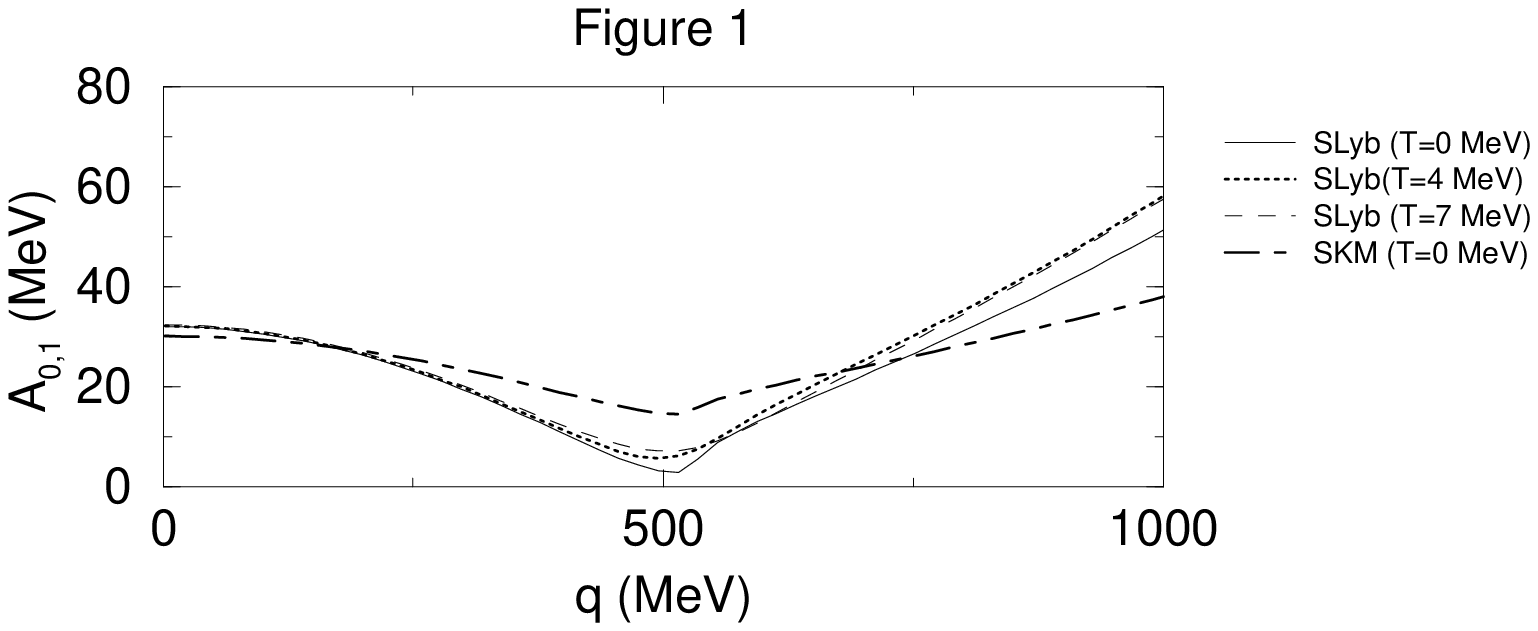,width=14cm} 
\end{figure}

\begin{figure}[htb]
\epsfig{figure=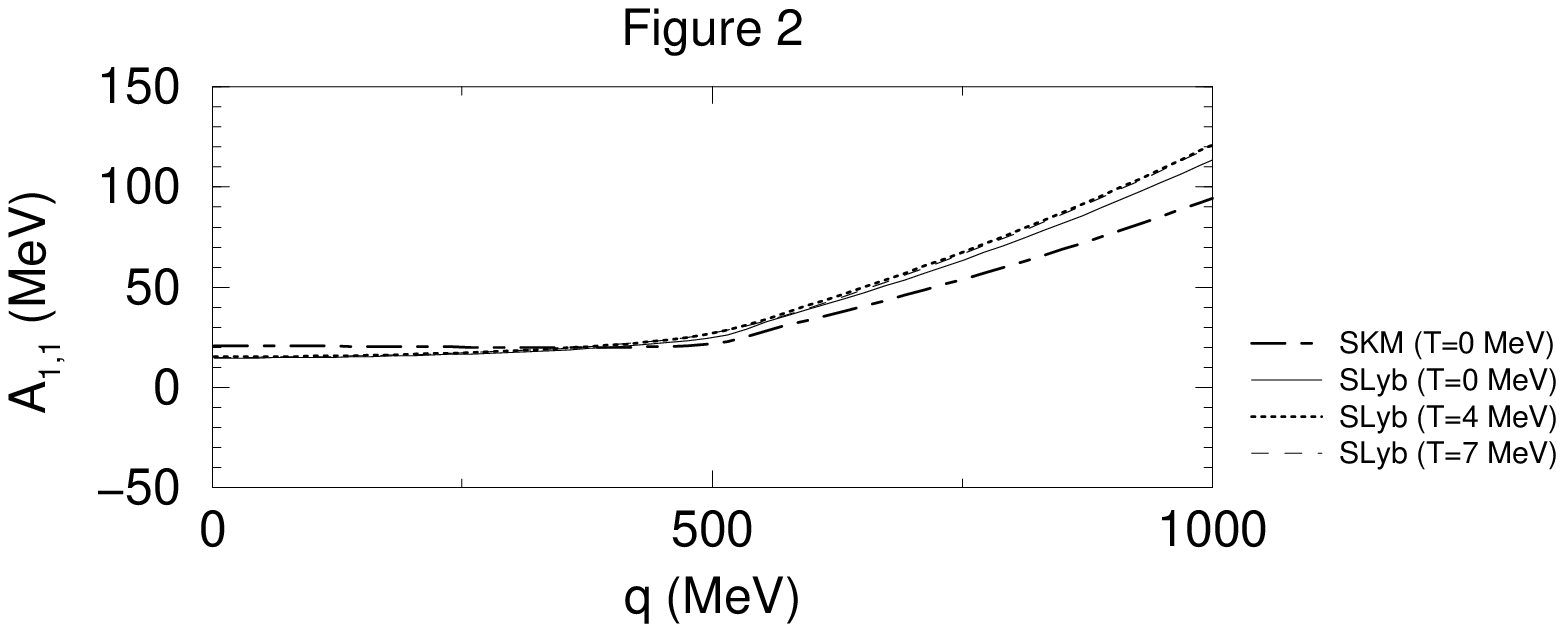,width=14cm} 
\end{figure}

\begin{figure}[htb]
\epsfig{figure=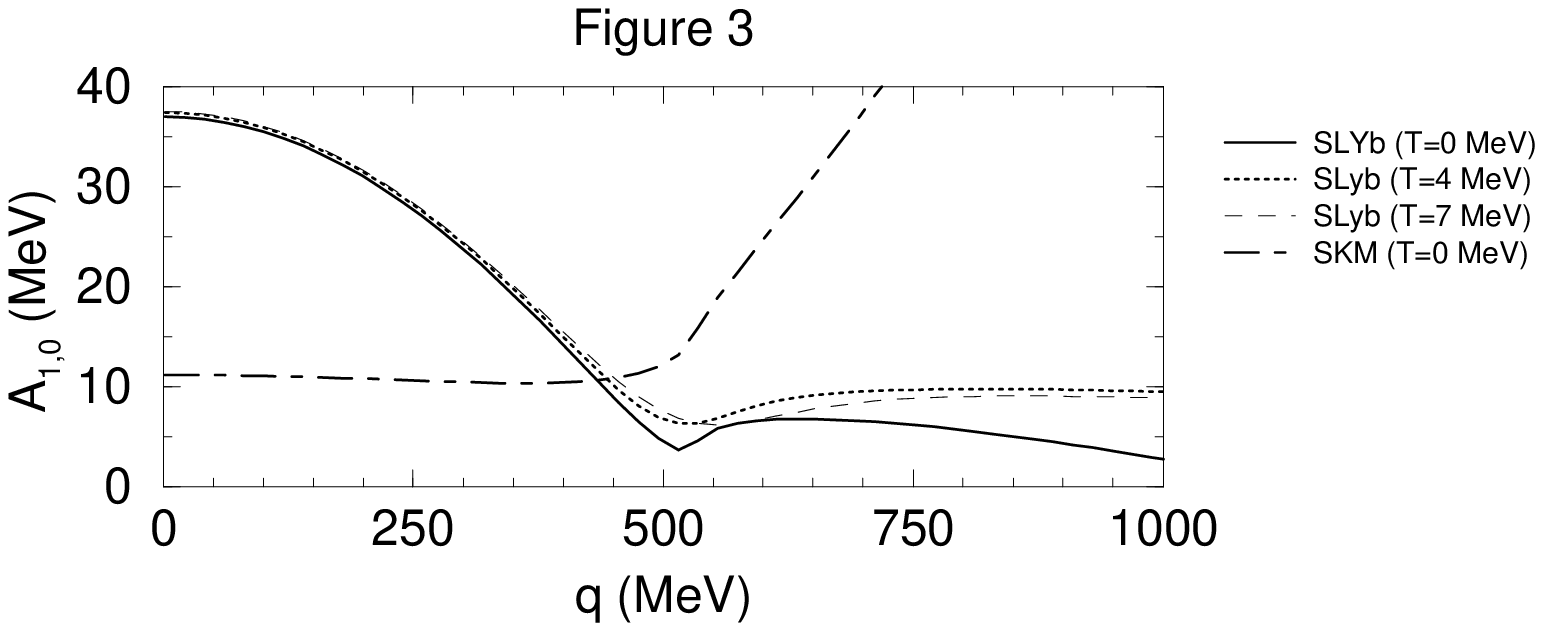,width=14cm} 
\end{figure}

\begin{figure}[htb]
\epsfig{figure=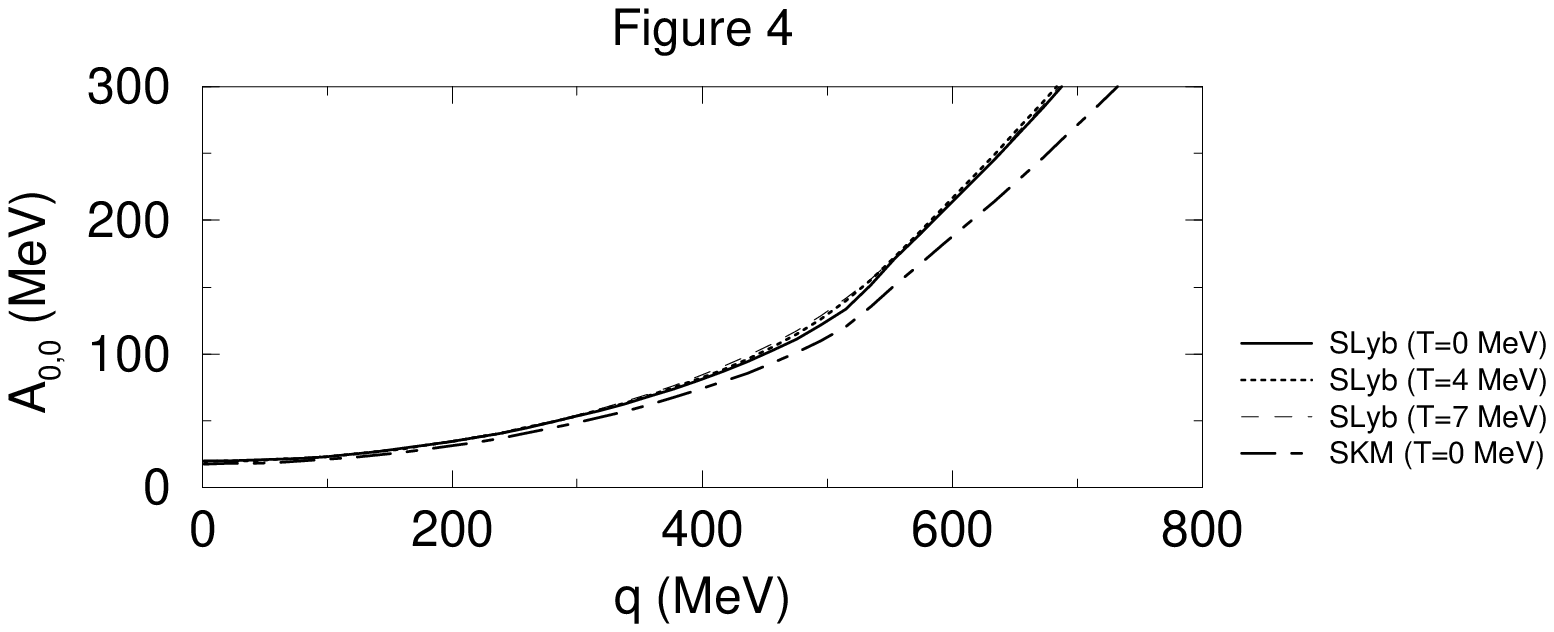,width=14cm} 
\end{figure}

\begin{figure}[htb]
\epsfig{figure=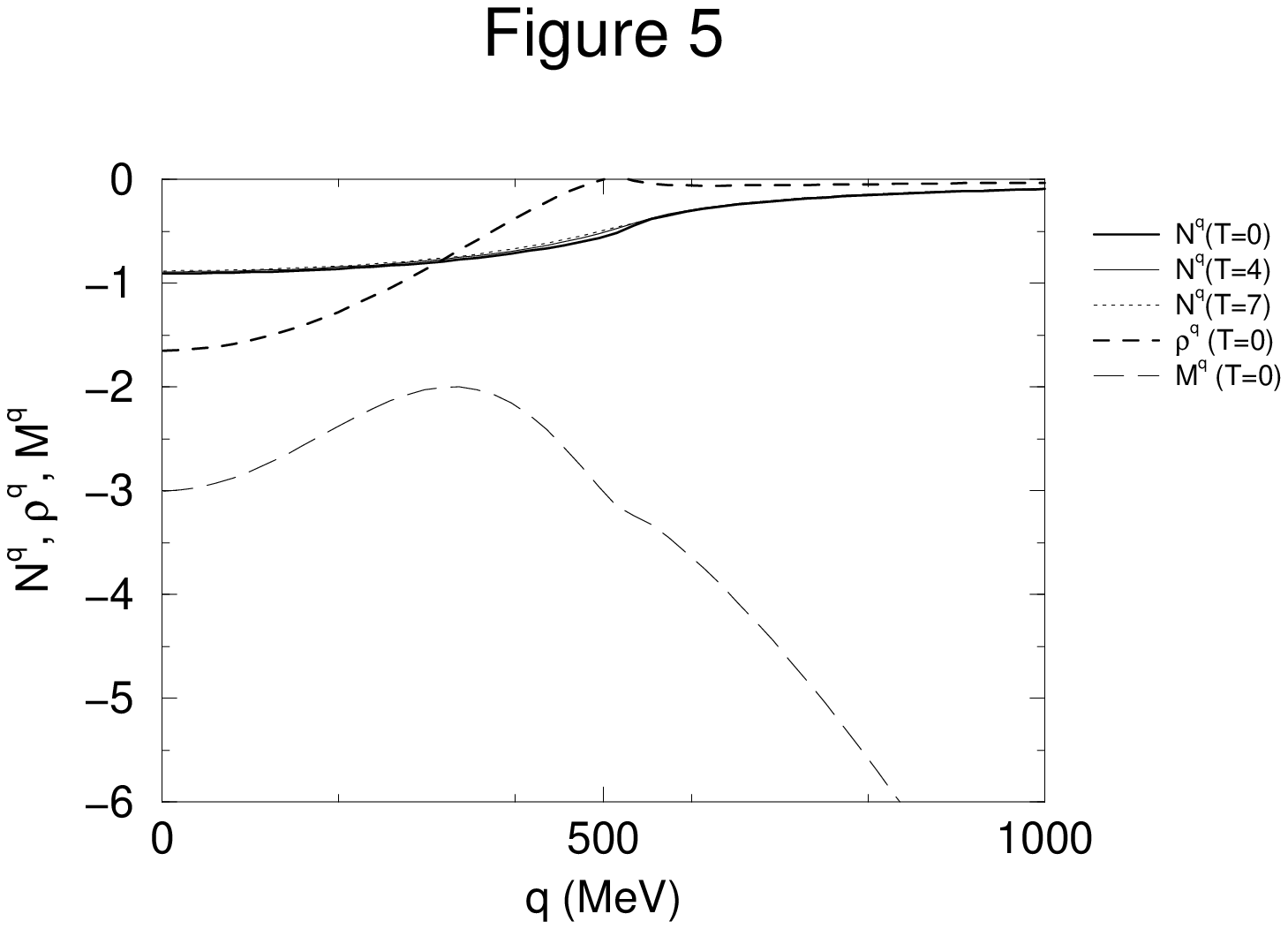,width=14cm} 
\end{figure}

\end{document}